# Role of Nano-fertilizer in Plants Nutrient Use Efficiency (NUE)

Mandana Mirbakhsh

*Department of Agronomy, Purdue University, West Lafayette. IN. 47907, USA.*

*Corresponding Author
Mandana Mirbakhsh, Department of Agronomy, Purdue University, West Lafayette. IN. 47907, USA.

**Abstract**
*Over the past half-century, the combination of technology and innovation has been developed to manage the negative impact of synthetic fertilizer on land ecosystems to identify the limitations of sustainability and optimize agricultural systems. The application of nano-fertilizers has achieved considerable interest due to their significant role as environmentally sustainable resource and soil health. Moreover, the increasing global population increased food insecurity especially under climate change. Nanotechnology has emerged as a promising alternative to help improving crop growth and productivity. However, un-balanced presence and long-term use of nano-particles alter photosynthesis, induce cellular redox imbalances resulting in lipid peroxidation, protein oxidation and DNA oxidative damage in plants, so more studies are still needed on their safe application with minimum side effects. In this paper, we reviewed nano-fertilizer mechanisms in plant, as well as their effects on microbiome and interaction with soil colloids. This study reviews the recent studies and findings about role of Nanotechnology in plant nutrition use efficiency to summarize a better understanding of this important issue. A comprehensive picture of ecological issues such as soil and water contamination in response to nano-fertilizer application will also be assessed.*

## 1. Introduction

The term "nano" has roots in Greek prefix meaning very small or "dwarf", which refers to small particle size ranges between 1 and 100 nm at least in one dimension that give credits to Nanoscience and Nanotechnology. However, Nanoscience and Nanotechnology differ from each other and should be distinguished. Nanotechnology applies the particles in devices and materials, but Nanoscience is the study of molecules on the scales of nanometers. Nanotechnology was not only the revolution in electronic and devices, but also considers as an innovative technique in agriculture resulting more nutrient efficiency, due to active penetration and high surface area of the particles. Employment of high-tech agricultural system with use of nanotools have a significant impact on fertilizer and pesticides improvement, which enhance quality and yield potential [1]. Nano-particles (NPs) have potential in triggering enzymes and releasing catalysts in plant-soil-microbial metabolism and interaction [2].

Nitrogen use efficiency (NUE) estimate the conversion of N inputs into agricultural products coupled with investigating N losses to environment [3]. NUE is calculated as a ratio of outputs over inputs that can estimates a metric range such as plant biomass per unit of N (e.g. fertilizer) applied [4]. NUE is a valuable indicator for monitoring sustainable development of food production and environmental impacts [5]. Slow and steady realizing of nano-materials not only reduces N losses and environmental contamination, but also enhance NUE [6].

The role of Nanotechnology in producing better and beneficial products in agriculture sector plays a significant role in increasing food availability and security by improving NUE [7]. Therefore, Nanotechnology has emerged as a promising alternative to end hunger, increase crop yield and soil health by enhancing the efficiency of agricultural inputs [8,9]. Nanotechnology revolutionized agriculture inputs including fertilizer, pesticide, sensors and transformed them into more advanced and efficient agricultural practices [10]. However, more research is needed to investigate the toxicity and degradation of nanomaterials in human body and ecosystem [11].

Nano-fertilizer are modified from traditional fertilizer, synthesized from fully bulked materials or extracted from different vegetative parts of plant by the help of different chemical and biological methods [12]. Since nano-fertilizers are tiny particles, they have higher surface area that facilitate their penetration, number of particles per unit area, and metabolic process [13]. Foliar application of nano-particles has shown the most significant yield increase [14]. Nano-fertilizer also recorded significant impact on early seed germination by penetrating into the seed and increase nutrient efficiency. Nanomaterials, such as ZnO, TiO2, MWCNTs, FeO, ZnFeCu-oxide are reported to increase crop growth and development with quality enhancement in many crop species including peanut, soybean, wheat, onion, spinach, tomato, potato and mustard in addition to germination [15,16]. Encapsulating fertilizer in nano-particles help them to increase their availability and prevent loss of nutrient and environmental contamination [17]. Although nano-fertilizer play great role in sustainable agriculture, crop production, and soil fertility, imbalanced implication of them causes soil organic matter (OM) loss and crop degradation. Comprehensive study is needed to investigate on economics and quality of nano-fertilizer and the most effective way for their application in field scale [18]. This paper discusses the nutrient use efficiency enhancement through nanotechnological interventions by reviewing the most recent studies in this field.

## 2. Recent Studies

In 2022, Sharma et al, studies the use and efficiency of nanotechnology in plant nutrient management [19]. The use of nano-technology in fertilizer industries promote higher production along with customization in nutrient values and improving higher nutrient use efficiency due to various physic-chemical properties. Nanoparticles easily delivered to the targeted cells in plant and rhizosphere [20]. Various environmental stress is alleviated by application of nano-fertilizer such as laboratory-prepared Mn, Cu, Zn, and Fe oxide in low concentration (<50 mg/L). The toxicity of these particles differed based on their molecular structures; although CuO and FeO recorded more toxicity than Cu and Fe ions, the toxicity of ZnO was similar to Zn [21,22]. Uncontrolled application of nano-particles is not a permanent alternative to increase productivity and alleviate hunger and limited food resources.

Reviewed plant nutrients absorbance and application of oxidized nano-particles and encapsulated-nanoparticles-fertilizer on plants [23]. Nutrient absorbance is due to the higher solubility of the nutrient released from fertilizer so plant uptake efficiency is low because of the conversion of these particles into unsolvable form in the field. Nano-fertilizer loaded on the bulk carrier acted as a substrate stop nutrient loss and enhance NUE by increasing the bioavailability of particles in soil [24]. Zinc nanoparticles boosts activity of enzymes, promote plant growth, and increase yield and biomass. Increasing sorghum yield and maize growth using zinc oxide nano-particles are the examples of successful application of Zn on plants [25].

In 2021, Mejias et al reviewed the potential use of nano-fertilizer as an innovative approach to reduce N losses and improve NUE by taking look at some consideration for animal food chain in grassland [26]. Since perennial crops have longer photosynthetic seasons that increase seasonal light interception and precipitation efficiencies, which can naturally increase NUE, a comprehensive analysis is needed on the potential use on N Nano-fertilizer on grassland [27-29]. Foliar application of Urea recorded beneficial and useful in grass covers but with high N volatilization losses [30]. More nano-fertilizer application on field scale is needed since most of the studies have only been carried out at a laboratory scale. Studying the adverse effects and time of application along with climate condition information is required in application of nano-fertilizer. Moreover, more investigation is needed to track nano-fertilizer transformation into ionic forms in the plant the protein, or some remain intact and reach the food chain of consumer.

Studied the vital roles of sustainable nano-fertilizers in improving plant quality and quantity in a review format which concluded that crop growth and nutrient use efficiency are improved by nano-fertilizer that reduce heavy metal toxicity and abiotic stress [31]. For example; salinization not only reduces soil health, quality, and crop yield, but also prevents foliar fertilizer's entrance due to saltiness of the nutrient ions [32-34]. It has been reported that nano-SiO2 (silicon dioxide) improves seed germination, photosynthesis rate, increases plant biomass with accumulation of proline in tomato and squash plants under salinity stress [35,36]. Torabian et al recorded the same result is in 2017 by foliar application of iron sulfate (FeSO4) in sunflower saline tolerance increment. Thereby, the regular fertilizer cannot compensate for crop and soil degradation, but the foliar nanoparticles can be transported from the application site to heterotrophic cells [37].

In 2019, Predoni et al examined the application of nanotechnology solutions in plant fertilization, which concluded the concentration of nanomaterials in water, soil or air is essential to determine their exposure to environment and health impact. The solubility of oxide nanoparticles has a great impact on cell structure responses, which have different pathway than metal toxicity in plants [38]. The concentration and rate of nanoparticles applications matters because their accumulation in plant biomass affect their fate and transport in the environment. Thereby, the process of utilization of nano-particles in agriculture should be monitored to avoid environmental contamination.

Reviewed and summarized recent attempts and usage of agro-nanotech innovation to face food problems faced in modern industrial agriculture [39]. 'Controlled loss fertilizer' is a new term that is used in agriculture to reduce non-point pollutions by forming a nano network by contacting with water and trapped fertilizer in soil network via hydrogen bonds or viscous force [40]. Their fixation into the soil, keep them around the crop roots and facilitate accessible easily absorption by plants in the time of need. In 2016,

Liu et al demonstrated 21.6% and 24.5% decreases in nitrogen runoff and leaching loss by control loss fertilizer application.

Biotic stressors such as pests and diseases have negative impact on crop productivity as well as abiotic stress. Nano-materials play significant role in minimizing yield losses and environmental stress. Strong antibacterial activity of Ag nano-particles on cotton plant, and controlling soil borne diseases caused by *Fusarium solani, Monilinia fructicola, Phytophthora infestans* by metal oxide nanomaterials such as CuO and MgO application in many plant species are some examples of studies on nano-materials-induced plant growth and biotic stress tolerance [41-44]. Ecofriendly nano-fertilizer not only decrease environmental contamination, but also improve soil fertility and microbial activity.

### 3. Ecological Issues with Nano-fertilizer

Although nano-materials considered as an alternative to alleviate food crisis and environmental stress, they can have a direct toxic effect on plant pathogens and/or can load chemicals that have antimicrobial effects [45]. Nanoparticle surface characteristics are acknowledged to be one of the most critical elements governing their stability and mobility as colloidal suspensions on adsorption, aggregation, and deposition. Interaction of nano-materials and cells through ligand is responsible for trafficking nano-fertilizer into the cells and any small alternation of their amount, size and charge is responsible for different interactions and change the balance [46]. Un-balanced presence of nano-particles alter photosynthesis and induced oxidative stress by ROS generation that resulted in DNA degradation, cell death, and Antioxidant enzymes inhibition [48]. Moreover, toxic metals and NPs can induce cellular redox imbalances and oxidative stress, resulting in lipid peroxidation, protein oxidation and DNA oxidative damage in plants [49]. It is also recorded that some of the nano-materials give more toxicity to the environment than others. For example; $TiO_2$, $SiO_2$ and ZnO with their role in mitigating biotic stress in crops recorded eliminating effect on $N_2$ fixation, which ends up to impaired growth of plants. It did not record any negative impact on seed germination, but the number of secondary roots decreased. Morphological change of bacterial cells also recorded by TiO2 [50]. ZnO can generate ROS on the surface of the particles and release zinc ions to inhibits the growth of mold and enhance crop production, but the responses are dose-dependent and can cause phytotoxicity to crops if not used correctly [51].

Empty silica NPs can be used as bio-stimulants, nano fertilizers, herbicides, and pesticides and act as nanocarriers for nucleotides, proteins, or other active molecules in agriculture [52]. Phytotoxicity and cytotoxicity in silver Nano-particles have been reported and the negative effects of silver Nano-particles such as AgNPs on plants' growth, microbial activity, and soil enzymes such as protease, catalase and peroxidase are confirmed. NPs' size, stability, and shape have a significant role in determining the toxicological effects of metal-based NPs on human health and ecosystems [53-56].

Carbon nanotubes (CNTs) is another class of nano-materials that have roles in crop production, abiotic stress alleviation and antimicrobial agents to plants. CNTs form when carbon atoms arrange in laminar structures in cylindric tube and few nanometers dimension. ROS generation has key role in in CNTs antimicrobial activity by formation of phytohormones such as jasmonic acid and salicylic acid in plant defense [57]. The accumulation and subsequent availability of these nanoparticles in food crops transfer of NPs into the food chain through edible plants is a great concern.

Copper oxide nano-particles (CuO NPs) is the other nano-particle that can pass easily through damaged root epidermis via apoplasty pathways and remains in intercellular space or roots for long time. CuO NPs inhibits root elongation much more than soluble Cu and it can explain why CuO NPs are more toxic than soluble Cu. CuO NPs can deposit in roots and leaf cell protoplasts and block Cu transportation to the shoot. The accumulated amount of Cu enters into cells by binding to protein carriers via ion channels or endocytosis and causes toxic impacts on the plant and inhibits its growth [58,59].

The mobility of nanoparticles depends on the rate of their infiltration or capture by stationary surfaces [60]. Surface charge is an important property that can determine the mobility of engineered nanomaterials in porous media [61]. Soil particles are normally negative and positively charged engineered nanomaterials will be electrostatically attracted to the soil surface [62]. Engineered nanomaterials having a negative charge will have more mobility due to stronger electrostatic repulsion between nanoparticles and soil and nanoparticles themselves. Therefore, various methods have been used to modify the surface properties of engineered nanomaterials to control the mobility of engineered nanomaterials in porous media.

The level of NPs in soil and water is increasing due to the growing consumer products that contained NPs on waste streams. Land runoff, precipitation, atmospheric deposition, and drainage are generally the main reasons for NPs pollutions. Nano-particles and micro pollutants are carried as runoff moves through lakes, rivers, wetlands, coastal water and ground water, which threaten human health [63]. Thereby, the leftover leachates from excess and after use of NPs ultimately find the way and accumulate in aggregates and colloids in their diverse types to the water bodies. These aggregates and colloids having NPs will generate additional anthropogenic waste (nano-waste) in the agroecosystem that is should be in continuous monitoring of the fate of nanoproducts leftover nano-waste products and soil composition, e.g., Arsenic is well-known groundwater contaminant and exists as both arsenate oxyanions at neutral pH under oxidizing conditions and arsenite under mildly reducing conditions at pH below 9.2, both can be remediated by using INP (iron nano-polymer) along with their corrosive products that tend to form aggregate during oxidative corrosion to iron oxide/hydroxide [64].

Nano-particles not only carried as runoff to the water, but also adsorb to soil and sediment particles very strongly due to their high surface areas [65]. When nano-particle absorbed to the immobile particle its mobility gets hindered, while sorption to the mobile particle may build up its mobility. The shape of the nanoparticle and collector surface has a considerable effect on the adhesion of nanoparticles to the surface [66]. The factor that controls the stability of nano-particles also tends to control the mobility properties of nanoparticles [67]. The difference between charges on nanoparticles and soil has an impact on electrostatic attraction or repulsion and aggregation among particles and between particles and soil [68]. All in all, application of nano-materials can improve plant growth, mitigate environmental stress, and decrease harmful effects caused by phytopathogens. However, nano-materials can have significant toxic impact on plants and environment in long-term commercial applications.

## 4. Conclusion

Nano-technology has emerged as a fantastic approach to improve plant health and resistance in case of environmental stress to provide food security, mitigate malnutrition with global population rise and climate change. The challenges in applying nano-materials to combat climate change and food insecurity can be highlighted as the need to study nano-materials structure, function, uptake and alternation in plant tissue [69]. Up to now, nano-materials played important role in agriculture and environmental science, but still more studies need to be done to clarify the security of their long-term application on field scale and their impact on soil microbiome, plants and human health [70].

As we reviewed, Nano-materials have been studied extensively on the experimental scale and the positive effects of nano-materials obviously observed in laboratory at low concentration in their short-term application. Antimicrobial effects and direct negative effect of nano-materials recorded on plant pathogens [71]. Although nano-materials can modulate metabolic pathways to alleviate biotic stress in plants more studies are still needed on their safe application with minimum side effects. Several factors such as size distribution, charge of the surface, treatment duration, and chemical features are involved to guarantee nano-materials safe application on plants, soil, and environments [72]. Moreover, Nano-materials accumulation in form of aggregate and colloids over causes soil and water contamination.

Importantly, another challenge is studying positive and negative consequences of slow rate of nano-particles degradation present in soil. Eco-friendly production of nano-fertiliser through the novel formulation of nano-technology, beside ecotoxicological studies of nano-materials are important factors that should be considered to favor public acceptance of this technology on a large scale in crop production [73,74].